\documentstyle[sprocl,epsf]{article}

\newcommand{\gapprox}{\raisebox{-.2ex}{$\stackrel{\textstyle>}
{\raisebox{-.6ex}[0ex][0ex]{$\sim$}}$}}
\newcommand{\lapprox}{\raisebox{-.2ex}{$\stackrel{\textstyle<}
{\raisebox{-.6ex}[0ex][0ex]{$\sim$}}$}}

\begin{document}

\begin{flushright} 
JLAB-THY-96-06 \\
September 1996
\end{flushright}

\vspace{0.5cm}

\title{QCD SUM RULES AND  VIRTUAL 
COMPTON SCATTERING\footnote{
Talk at the Workshop on Virtual Compton Scattering,
Clermont-Ferrand, France, June 26-29,1996.}}

\author{A.V. RADYUSHKIN\footnote{Also Laboratory
of Theoretical Physics, JINR, Dubna, Russian Federation}}

\address{Physics Department, Old Dominion University,
\\ Norfolk, VA 23529, USA 
 \\ and  \\
Thomas Jefferson National Accelerator Facility \\
 Newport News, VA 23606, USA}

\maketitle\abstracts{In this talk  I  report on   recent progress in 
a few areas closely related to the virtual Compton scattering studies.
In particular,  I  discuss the quark-hadron duality 
estimate of the $\gamma^* p \to \Delta^+$
transition, QCD sum rule calculation 
of the $\gamma \gamma^* \to \pi^0$ form factor,  
 and application of perturbative QCD to
deeply virtual Compton scattering at small $t$. 
}

\section{Soft $vs.$ hard dynamics in QCD }

{\it QCD and virtual Compton scattering.} The  kinematics of the 
amplitude of the process $\gamma^* p \to \gamma  p'$ can be 
specified  
by the initial nucleon momentum $p$,
the   momentum  transfer $r=p-p'$ and  the 
momentum $q$  of the initial virtual  photon, $q^2 \equiv -Q^2$.
The final  photon momentum $q'$ is then given by
$q'= q +r$ with  $q'^2=0$.  Other important momentum invariants are
$t \equiv (p'-p)^2=(q-q')^2$ and $s \equiv (q+p)^2$.

Taking  $Q^2$ large, $i.e.,$ at least above $1 \, GeV^2$,
 one can hope  to enter the region where 
the  amplitude  is dominated by short distances 
between the two photon vertices and pQCD may be applicable.
In this situation,   it is tempting to speak about the
``virtual Compton scattering on a single quark'' implying that 
the large-$Q^2$ behaviour is given just by 
the quark propagator  (see Fig.$1a$), 
while the long-distance information 
is accumulated in a distribution function $F(X,t)$ described by 
the matrix element of 
$\langle p' | \bar q \ldots q | p \rangle$ type. 
However,  the  factorization formula
\begin{equation}
M(Q^2,s,t) \to  \int  m(s/Q^2,X) F(X,t) dX
\label{fact}
\end{equation} 
only makes sense if  $|t| \ll Q^2$.
Otherwise,  if  $|t| \sim Q^2$,  large momentum
enters into the hadron wave function and one deals with
the scattering  process on the hadron as a whole. 

\begin{figure}[htb]
\mbox{
   \epsfxsize=8cm
 \epsfysize=3cm
 \hspace{2.0cm}  
 \epsffile{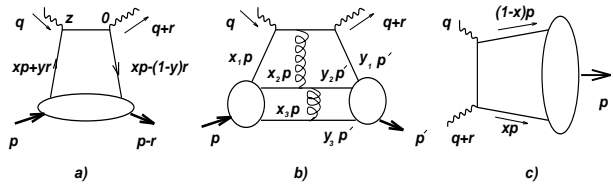} 
 }
  \vspace{-0.5cm}
{\caption{\label{fig:1}
$a)$ Handbag diagram contributing 
 to the DVCS amplitude at  $t=0$. The lower blob corresponds 
to  the double quark distribution $F(x,y)$.
$b)$ One of the two-gluon-exchange diagrams dominating 
the VCS amplitude for asymptotically 
large $Q^2$ and $t$. $c)$ 
 Lowest-order diagram for the $\gamma^* \gamma \to \pi^0$
transition form factor. The  blob corresponds 
to  the pion  distribution amplitude 
$\varphi_{\pi}(x)$. 
  }
}
\end{figure}

For $|t| \ll Q^2$,  the function $F(X,t)$  in  eq.(\ref{fact}) looks  
like a parton distribution function $f(x)$ with  an additional
form-factor-type  dependence on $t$.
To make analogy with deep inelastic scattering,
it  is instructive to recall that  the 
imaginary part of the virtual forward Compton 
amplitude (for which $q'=q$ and $p'=p$)  in the 
 limit of large $Q^2$ and fixed  Bjorken
variable $\zeta \equiv Q^2/2(pq)$
can be written  as 
\begin{eqnarray}
\int_0^1 f(x) \delta((q+xp)^2) \cdot 2 (qp) dx 
= \int_0^1 f(x) \delta(x-Q^2/ 2 (qp) )  dx  = f(\zeta)\, , 
\end{eqnarray} 
where $xp$ is the fraction of the initial hadron momentum carried by 
the interacting quark.  The  usual parton distribution 
functions $f(\zeta)$  correspond to  exactly forward matrix elements,
with $r \equiv  p' -p =0$,
while the kinematics of  VCS requires that $r \neq 0$ and 
$t  \equiv r^2 \neq 0$.  Hence,  we need a new type of parton 
functions $F(X,t)$ \cite{ji}. In the limit $t \to 0$,
they reduce to the ``asymmetric distribution functions''   
$F(X)$ \cite{compton,gluon}.
Hence, the studies of deeply virtual Compton scattering
(DVCS)  are related to a new field of pQCD applications.
As shown in refs.\cite{compton,gluon},
the asymmetric distributions $F(X)$ have features  
of both the distribution functions and 
distribution  amplitudes (wave functions).
A more detailed discussion of DVCS at small $t$
will be given in Section 3.

Another situation in which pQCD is applicable 
is when both $Q^2$ and $|t|$ are asymptotically  large.
Then the virtual Compton scattering amplitude
factorizes into a convolution of the short-distance amplitude
$m(\{x_i\},\{y_j\},Q^2,s,t)$ and two 
distribution amplitudes $\varphi(x_1,x_2,x_3)$,
$\varphi(y_1,y_2,y_3)$ describing the proton in the initial 
and final state, respectively (see Fig.$1b$).  They 
are  related to matrix elements
of $\langle 0|q\ldots q \ldots q |p \rangle$ type and  
give  the probability amplitude, 
that  $e.g.,$ the initial proton can be treated as 
three collinear quarks with the momentum $p$ divided into
fractions $x_1 p,x_2 p, x_3 p$ with $x_1+x_2+x_3 =1$.
The short-distance amplitude $m(\{x_i\},\{y_j\},Q^2,s,t)$
is given by Feynman diagrams involving two hard gluon exchanges,
suppressed by a  factor 
$(\alpha_s/\pi)^2 \sim 1/100$ compared to the ``soft contribution''
produced  by a simple overlap of soft wave functions, without 
any gluon exchanges.
The soft term, however, has an extra power
of $1/Q^2$ for large $Q^2$. As a result, the hard term asymptotically
dominates, though  the soft term may be 
much larger than the  hard one for accessible $Q^2$.

{\it Quark-hadron duality and $\gamma^* p \to \Delta^+$ transition.}
In particular, a purely soft contribution to $G_M^p(Q^2)$ 
calculated within the local quark-hadron duality
approach \cite{nr83} is in good agreement with 
experimental data up to $Q^2 \sim \, 20 \, GeV^2$.
The same approach was used recently \cite{del} to get 
the estimates of  the soft term  for the 
$\gamma^* p \to \Delta^+$ transition.
For  the   magnetic form 
factor $G_M^*(Q^2)$ these estimates are rather
close to the results of the analysis of inclusive
SLAC data \cite{stoler,keppel}.
A small value for the ratio $G_E^*(Q^2)/G_M^*(Q^2)$
obtained  in ref.\cite{del} is also in agreement with available 
data \cite{burkert}, in contrast to the pQCD prediction
\cite{carlson} 
which gives $G_E^*(Q^2)/G_M^*(Q^2) \to \, -1$ 
for the ratio of hard contributions. 
\begin{figure}[htb]
\epsfxsize=6cm
\epsfysize=4cm
 \hspace{-0.5cm}
\epsffile{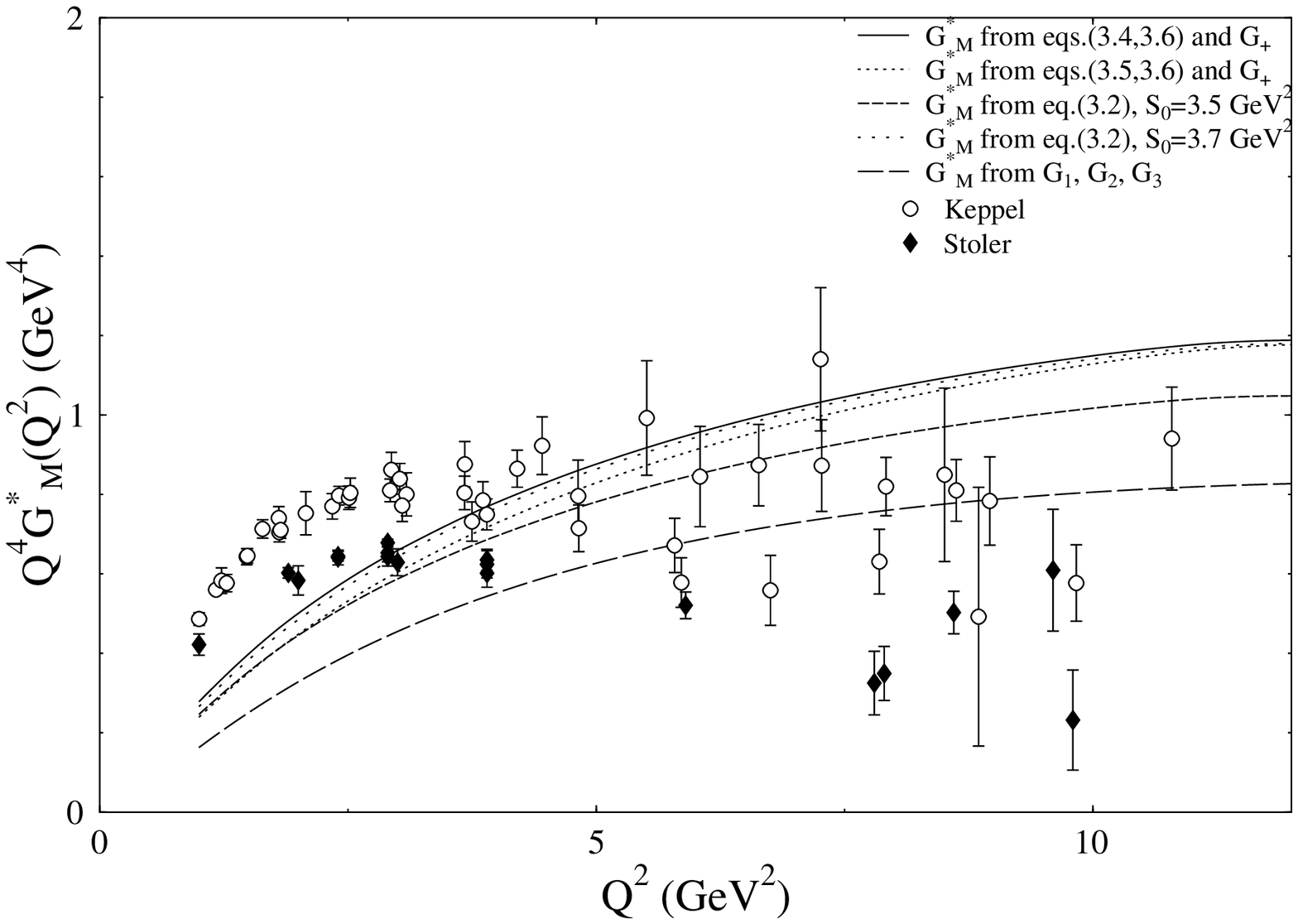}
\epsfxsize=6cm
\epsfysize=4cm
\hspace{0.1cm}
\epsffile{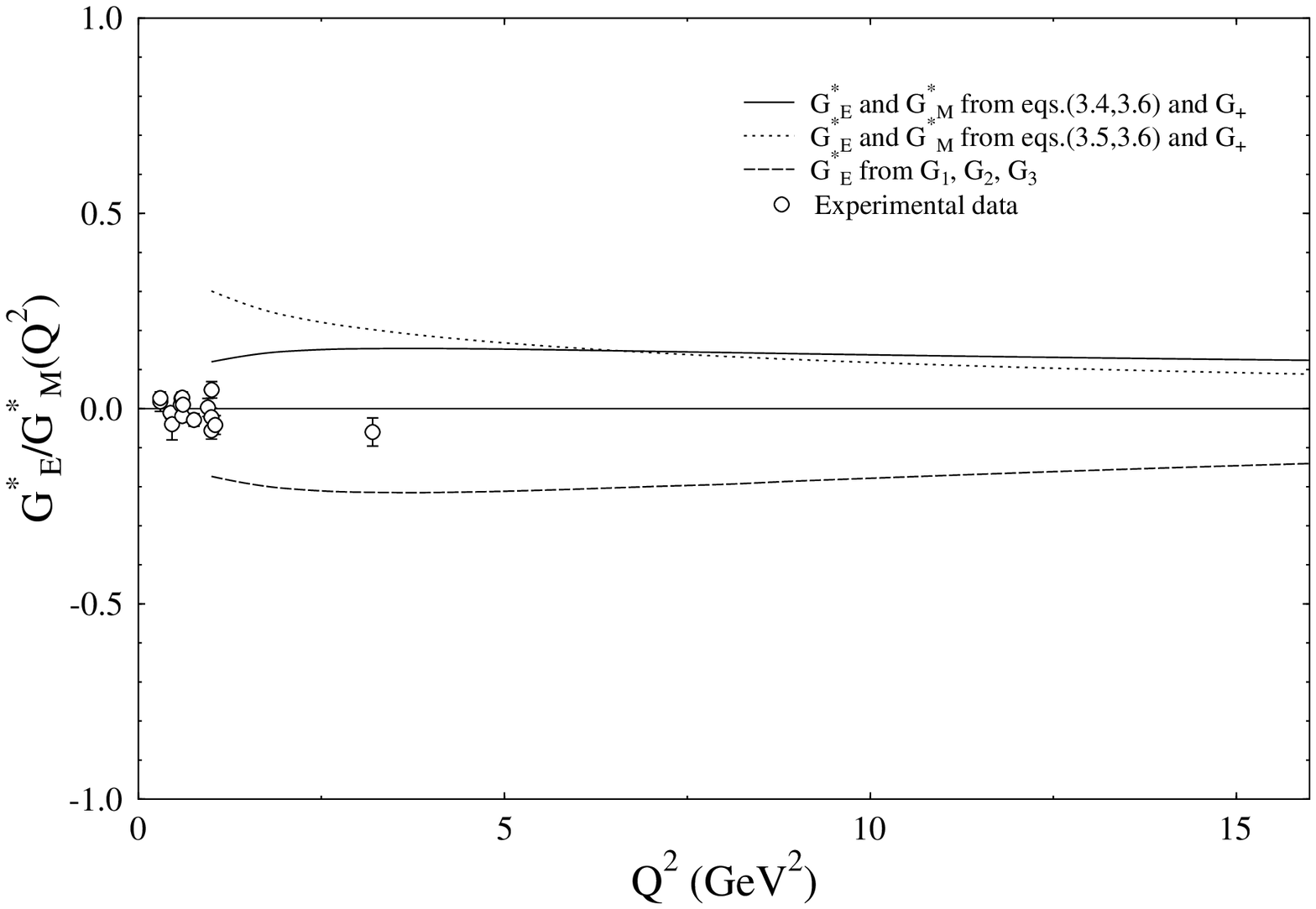}
 \vspace{-1cm}
 {\caption{\label{fig:2} Quark-hadron duality estimates
for the $\gamma^* p \to \Delta^+$ transition.
 Left: form factor $G_M^*(Q^2)$. Right: 
  ratio of form factors $G_E^*(Q^2)$ and $G_M^*(Q^2)$. }}
\end{figure}

Within different nonperturbative approaches
\cite{nr82,nr83,chizhit,kroll96},
  it was observed that
 soft terms are  sufficiently large to describe the data 
or that the hard terms are too small compared to the data.
Hence,  there is a growing evidence 
that  soft terms dominate the exclusive amplitudes
at accessible energies.
Of course, the magnitude of the hard 
contribution depends on the shape of   distribution amplitudes (DA's).
The latter  are usually  
integrated with the weights like $1/x_1  x_2$,
so the humpy DA's of Chernyak-Zhitnitsky (CZ) type \cite{cz82,cz84}
produce  contributions which are much larger than those
obtained with smooth DA's close to the ``asymptotic'' forms.
The CZ wave functions were originally motivated
by QCD sum rule analysis \cite{cz82}. 
 However,  the results of the QCD sum rule 
calculations of the  DA's are extremely model-dependent and unreliable. 
 Furthermore, for  the theoretically most 
clean  case  of the $\gamma^* \gamma \to \pi^0$ form factor,
both a direct QCD sum rule
calculation of this form factor 
and available experimental data show 
no enhancement compared to the  pQCD result  obtained with the 
asymptotic DA for the pion.

\section{$\gamma^* \gamma \to \pi^0$  form factor}

{\it   pQCD analysis.} 
The transition $\gamma^* \gamma^* \to \pi^0$
of two virtual photons $\gamma^*$ into a neutral pion 
provides an exceptional opportunity to test 
QCD predictions for exclusive processes.  
In the lowest order of 
perturbative QCD, its asymptotic behaviour 
is due to the subprocess 
$\gamma^*(q) + \gamma^*(q^{\prime}) \to \bar q(\bar xp) + q (xp) $
with $x$ ($\bar x$) being  the fraction of the pion momentum $p$ carried
by the quark produced at the $q$ ($q')$ photon vertex (see Fig.1$c$).
The relevant diagram is similar to the handbag
diagram for deep inelastic scattering,
with the main difference that one should use
 the pion distribution amplitude $\varphi_{\pi}(x)$
instead of parton densities.
For large $Q^2$, the perturbative 
QCD prediction is given by \cite{bl80}:
\begin{equation}
F_{\gamma^* \gamma^*  \pi^0 }^{as}(Q^2, q^{\prime 2}) = \frac{4\pi}{3}
\int_0^1 {{\varphi_{\pi}(x)}\over{xQ^2 - \bar x q^{\prime 2} }} \, dx 
\stackrel{q^{\prime 2}=0}{\longrightarrow}
\frac{4\pi}{3}
\int_0^1 {{\varphi_{\pi}(x)}\over{xQ^2}} \, dx
\equiv \frac{4\pi}{3Q^2} I \, .
\label{eq:gg*pipqcd}
\end{equation}
Experimentally, 
the most important situation is when the lower virtuality photon
is (almost) real $q^{\prime 2} \approx 0$. In this  case,  necessary 
nonperturbative information  
is accumulated   in 
the same integral $I$ (see  eq.(\ref{eq:gg*pipqcd}))
that appears  in the one-gluon-exchange 
diagram  for the  pion electromagnetic  form factor  
\cite{cz84,pl80,blpi79}.
The  value of $I$ depends on the shape of the 
pion distribution amplitude $\varphi_{\pi}(x)$.
In particular,  using  the 
asymptotic form 
$
\varphi_{\pi}^{as}(x) = 6 f_{\pi} x \bar x   
$
\cite{pl80,blpi79} gives $F_{\gamma \gamma^*  \pi^0 }^{as}(Q^2) = 
4 \pi f_{\pi}/Q^2 $ for  the asymptotic
behaviour \cite{bl80}. If one takes the
Chernyak-Zhitnitsky form  \cite{cz82}
$\varphi_{\pi}^{CZ}(x) = 30 f_{\pi} x(1-x)(1-2x)^2$,
the integral $I$ increases by a sizable factor of 5/3, 
and this difference can be used for experimental
discrimination between the two forms.

Note, that
the pQCD  hard scattering term for $\gamma \gamma^* \to \pi^0$ 
 has the zeroth 
order in the QCD coupling constant $\alpha_s$,
just like in deep inelastic
scattering. Hence, there are good reasons to  
expect that pQCD   for $F_{\gamma \gamma^*  \pi^0 }(Q^2)$ 
 may work at rather low  $Q^2$. 
The $Q^2=0$ limit of 
$F_{\gamma \gamma^*  \pi^0 }(Q^2)$ 
is  known from  $\pi^0 \to \gamma \gamma$ decay rate. 
Using  PCAC and ABJ anomaly \cite{ABJ},
one can calculate $F_{\gamma \gamma^*  \pi^0 }(0)$
theoretically:
$
F_{\gamma \gamma^*  \pi^0 }(0) =1/ \pi f_{\pi} .
$
It is natural to expect that a complete QCD result 
does not  strongly deviate from 
a simple interpolation  \cite{blin}
$
\pi f_{\pi} F_{\gamma \gamma^*  \pi^0 }(Q^2) = 
1/(1+ Q^2/4 \pi^2 f_{\pi}^2)  
$ 
between 
the $Q^2=0$ value and the large-$Q^2$ 
asymptotics \footnote{In particular, such an 
interpolation agrees with the results of  a 
constituent quark model calculation \cite{hiroshi}}.  
This interpolation implies the asymptotic form 
of the distribution amplitude for the large-$Q^2$ limit and 
agrees  with CELLO
experimental data \cite{CELLO}.
It was also claimed \cite{CLEO} that the new CLEO data  
available  up to $8 \, GeV^2$  also
agree with the interpolation  formula.

Comparing the data with  theoretical predictions,
one should take into account 
the one-loop pQCD radiative corrections to the hard 
scattering amplitude  calculated in ref.\cite{braaten}.
Effectively,  the correction 
decreases the leading-order result by 
about  $20 \%$, still leaving a sizable 
gap between the prediction based on the CZ amplitude
and the phenomenologically
successful Brodsky-Lepage interpolation formula\cite{blin}.  
Hence, the new preliminary   data\cite{CLEO} seem to  
indicate that the magnitude of $I$
is close to that corresponding to the asymptotic 
form of the pion distribution amplitude.
Because of the far-reaching consequences   of this conclusion,
it is desirable  to have  a direct   calculation
of  the  $\gamma\gamma^* \to \pi^0$ 
form factor in the intermediate region of  moderately large 
momentum transfers $Q^2 \gapprox 1 \, GeV^2$. 
 As we will see below, 
 the QCD sum  rules   allow one  to 
calculate $F_{\gamma \gamma^*  \pi^0 }(Q^2)$  for large $Q^2$
without any  assumptions
about the shape of the pion distribution amplitude,
and the result 
can be used to get  information about 
 $\varphi_{\pi}(x)$.

{\it QCD sum rules and pion distribution amplitude.}
The CZ sum rule written directly for the pion distribution amplitude
$\varphi_\pi(x)$ is 
\begin{eqnarray}
f_\pi\varphi_\pi(x)&=&\frac{3M^2}{2\pi^2}(1-e^{-s_0/M^2})x(1-x)
  +\frac{\alpha_s\langle GG\rangle}{24\pi M^2}[\delta(x)+\delta(1-x)]
\nonumber \\
		&+ &  \frac{8}{81}\frac{\pi\alpha_s\langle\bar
		   qq\rangle^2}{M^4}
\{11[\delta(x)+\delta(1-x)]+2[\delta^\prime(x)+\delta^\prime(1-x)]\}.
\label{eq:wfsr}
\end{eqnarray}
Here, $M$ is the auxiliary Borel  parameter which must be 
taken in the region
where the r.h.s. is least sensitive to 
its variations, and $s_0$ is the effective
onset of the continuum  fitted to maximize the $M^2$-stability region.
From the QCD sum rule for $f_{\pi}$, \cite{svz}
 $s_0 \approx 0.7 \, GeV^2$.
 As emphasized in ref.\cite{MR},   
 the lowest condensates  $\langle GG\rangle$ and 
$\langle \bar qq\rangle^2$ taken into account  in eq.(\ref{eq:wfsr})
 do not provide all the information necessary
for a reliable determination of  $\varphi_\pi(x)$.
The   humpy CZ shape is, in fact, a compromise 
between the $\delta(x)$,  $\delta(1-x)$ condensate 
 peaks and the smooth $x(1-x)$ 
behaviour of the perturbative term.
Adding higher condensates, $e.g.,$ $\langle \bar q D^2 q\rangle$,
one would get even higher derivatives of
$\delta(x)$ and $\delta(1-x)$.
The sum  of such singular terms
can be treated as  an expansion
of some finite-width function $\delta \varphi (x)$:
\begin{equation}
\delta \varphi (x) = a_0 \delta(x) + a_1 \delta^{\prime} (x) + a_2 \delta^{\prime \prime}(x)  + \ldots 
+ \{x \to 1-x \}.
\end{equation}
Of course, the knowledge of $a_0$ alone is not sufficient for a reliable 
reconstruction of $\delta \varphi (x)$.
On the other hand, the higher coefficients $a_1, a_2, etc.$
are given by a sum of several higher condensates whose magnitudes
are completely unknown. Hence, no strict conclusions can be made.
The CZ procedure is equivalent to assuming that $a_1, a_2,  \ldots \, \sim  0$,
though other choices ($e.g.,$  nonlocal condensate model \cite{MR})
may look more realistic.

{\it QCD sum rule for doubly virtual form factor
$F_{\gamma^* \gamma^* \pi^0} (q^2, q^{\prime 2})$.}
Instead of following the steps dictated by the old logic:  
$1)$  pQCD factorization for $F_{\gamma^* \gamma \pi^0} (Q^2)$;
$2)$ QCD sum rules for the moments of  
  $\varphi_{\pi} (x)$ (which are unreliable);
$3)$ calculation of $I= \int_0^1 \varphi_{\pi} (x)/x \, dx$, 
 we  developed in ref. \cite{rr}  the approach which  
$1)$ starts with the  QCD sum rule  for $
F_{\gamma^* \gamma^* \pi^0} (q^2, q^{\prime 2})$
in the $q^{\prime 2} \to  0$ limit;
$2)$   information about $I$ is extracted from this sum rule and
$3)$ then  used to make conclusions 
about the shape of  $\varphi_{\pi} (x)$.
When  both  
virtualities of the photons are large, we have the following QCD sum rule:
\begin{eqnarray}
\pi f_{\pi} \mbox{$F_{\gamma^*\gamma^*\pi^\circ}$}(Q^2, q^{\prime 2})=
 2\int_0^{s_o} ds \, e^{-s/{M^2}}  
\int_0^1 \frac{x\bar{x}(xQ^2 - \bar x q^{\prime 2})^2}
{[s{x}\bar{x}+xQ^2 - \bar x q^{\prime 2}]^3} \,dx  \,
\nonumber \\  
+\frac{\pi^2}{9}
{\langle \frac{\alpha_s}{\pi}GG \rangle}
\left(\frac{1}{2M^2 Q^2} - \frac{1}{2M^2 q^{\prime 2}} 
 + \frac1{Q^2 q^{\prime 2}}\right) 
 \nonumber\\
+ \frac{64}{243}\pi^3\alpha_s{\langle \bar{q}q\rangle}^2
\left( \frac1{M^4} 
\left [ \frac{Q^2}{q^{\prime 4}} - 
\frac9{2q^{\prime 2}}+\frac9{2Q^2}-\frac{q^{\prime 2}}{Q^4} \right ] + 
\frac9{Q^2 q^{\prime 4}} -\frac9{Q^4q^{\prime 2} } \right )  .
\label{eq:SR1}
\end{eqnarray}
In this  situation, the  pQCD approach
is also expected to work. 
Indeed,  neglecting  the $s{x}\bar{x}$-term 
compared to $xQ^2 - \bar x q^{\prime 2}$ 
and keeping only the leading  $O(1/Q^2)$
and $O(1/q^{\prime 2})$ terms in the condensates, 
 we can write eq.(\ref{eq:SR1}) as
\begin{eqnarray}
\mbox{$F_{\gamma^*\gamma^*\pi^\circ}$}(,Q^2) 
= \frac{4\pi}{3f_{\pi}}
   \int_0^1 \frac{dx}{  xQ^2 - \bar x q^{\prime 2}} \, 
\left \{ \frac{3M^2}{2\pi^2}(1-e^{-s_0/M^2}) x\bar{x} 
\right. \nonumber \\ \left.
+ \frac{1}{24M^2}
\langle \frac{\alpha_s}{\pi}GG\rangle [\delta(x) + \delta (\bar{x})] 
\right. \nonumber \\
+ \left. \frac{8}{81M^4}\pi\alpha_s{\langle \bar{q}q\rangle}^2
 \biggl ( 11[\delta(x) + \delta (\bar{x})] +  
2[\delta^{\prime}(x) + \delta ^{\prime}(\bar{x})] 
\biggr ) \right \} 
\label{eq:SRlargeQ2wf}. 
\end{eqnarray}
The expression in  curly brackets 
coincides with the QCD sum rule (\ref{eq:wfsr})
for 
the pion distribution amplitude 
$f_{\pi} \varphi_{\pi}(x)$.   
Hence, when both $Q^2$ and  $q^{\prime 2}$ are 
large,  
the QCD sum rule
(\ref{eq:SR1})
exactly reproduces  the pQCD result (\ref{eq:gg*pipqcd}).

One may be tempted to  get 
a QCD sum rule for the integral $I$ by taking  $ q^{\prime 2}=0$ 
in eq.(\ref{eq:SR1}).
Such an attempt, however, fails immediately 
because of the power singularities 
$1/q^{\prime 2}$, $1/ q^{\prime 4}$, $etc.$
in the condensate terms.
It is easy to see that these singularities 
are produced by  the 
$\delta(x)$ and $\delta'(x)$ terms in eq.(\ref{eq:SRlargeQ2wf}).
In fact, it is precisely these terms that  generate the two-hump form 
for $\varphi_{\pi}(x)$ in the CZ-approach \cite{cz82}. 
 The advantage of having  a direct sum rule for 
$F_{\gamma \gamma^* \pi^0}(Q^2,q^{\prime 2})$ 
is that the small-$q^{\prime 2}$ behavior of
$F_{\gamma \gamma^* \pi^0}(Q^2,q^{\prime 2})$
is  determined by the position of the closest  resonances 
in the $q^{\prime }$ channel, which is known.
Eventually,  $1/q^{\prime 2}$  
is  substituted  for small $q^{\prime 2}$ by something
like $1/m_{\rho}^2$ and the QCD sum rule in the 
 $q^{\prime 2}$  limit is 
\begin{eqnarray}
& \,& \pi f_{\pi} F_{\gamma \gamma^* \pi^0}(Q^2) =
\int_0^{s_0}
\left \{ 
1 - 2 \frac{Q^2-2s}{(s+Q^2)^2} 
\left (s_{\rho} - \frac{s_{\rho}^2}{2 m_{\rho}^2} \right ) 
\right.  \nonumber \\ 
&+& \left. 2\frac{Q^2-6s+3s^2/Q^2}{(s+Q^2)^4} \left (\frac{s_{\rho}^2}{2}
 - \frac{s_{\rho}^3}{3  m_{\rho}^2} \right ) 
\right \} 
 e^{-s/M^2} 
\frac{Q^2 ds }{(s+Q^2)^2} 
 \nonumber \\ 
&+&\frac{\pi^2}{9}
{\langle \frac{\alpha_s}{\pi}GG \rangle} 
\left \{ 
\frac{1}{2 Q^2 M^2} + \frac{1}{Q^4} 
- 2 \int_0^{s_0} e^{-s/M^2} \frac{ds }{(s+Q^2)^3} 
\right \} 
 \nonumber \\ 
&+&\frac{64}{27}\pi^3\alpha_s{\langle \bar{q}q\rangle}^2
\lim_{\lambda^2 \to 0}
\left \{ 
\frac1{2Q^2 M^4} 
+ \frac{12}{Q^4 m_{\rho}^2 } 
\left [ 
\log \frac{Q^2}{\lambda ^2} -2 
\right.  \right. \nonumber \\
&+& \left.  \left. \int_0^{s_0} e^{-s/M^2} 
\left ( 
\frac{s^2+3sQ^2+4Q^4} {(s+Q^2)^3} - \frac1{s+\lambda ^2} 
\right) ds 
\right] 
\right. 
 \label{eq:finsr} \\ 
&-& 
\left.  
\frac4{Q^6}
\left [ 
\log \frac{Q^2}{\lambda^2} -3+
\int_0^{s_0} e^{-s/M^2} 
\left (  
\frac{s^2+3sQ^2+6Q^4} {(s+Q^2)^3} - \frac1{s+\lambda ^2}
\right) ds 
\right] 
\right \} .
\nonumber
\end{eqnarray}

\begin{figure}[htb]
\mbox{
\hspace{1.5cm}
\epsfxsize=5cm
 \epsfysize=7.5cm
 \epsffile{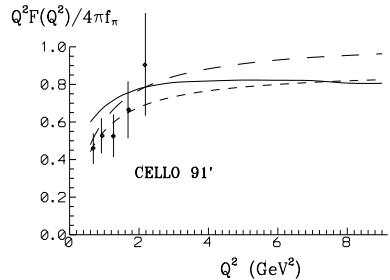} 
}
  \vspace{-3.5cm}
{\caption{\label{fig:3}
 Combination  $Q^2 F_{\gamma \gamma^* \pi^0}(Q^2)/4\pi f_{\pi}$ 
as calculated from the QCD sum rule 
(solid line), $\rho$-pole model (short-dashed line) 
and Brodsky-Lepage interpolation (long-dashed line).
 }}
\end{figure}

In Fig.\ref{fig:3}, we present a curve for 
$Q^2F_{\gamma \gamma^* \pi^0}(Q^2)/4\pi f_{\pi}$ 
calculated from eq.(\ref{eq:finsr}) for 
standard values of  the condensates,
 $\rho$- and $\pi$-meson duality intervals 
 $s_{\rho} = 1.5 \, GeV^2$,  \cite{svz},
$s_0 = 0.7 \, GeV^2$
and $M^2 = 0.8\, GeV^2$. 
It is rather close to the curve corresponding to the 
Brodsky-Lepage interpolation
formula  $\pi f_{\pi} F_{\gamma \gamma^* \pi^0}(Q^2) = 
1/(1+Q^2/4\pi^2 f_{\pi}^2)$
and to that based on the $\rho$-pole  approximation 
$\pi f_{\pi} F(Q^2) = 1/(1+Q^2/m_{\rho}^2)$.
Hence,  our result favors 
a  pion  distribution amplitude 
which is close to  the asymptotic form.
It should be noted,
that  the 
$\rho$-pole behaviour in the $Q^2$-channel  has nothing  to do 
with the explicit use of the $\rho$-contributions
in our models for the correlators in the $q^{\prime 2}$-channel:
the  $Q^2$-dependence of the $\rho$-pole type  emerges 
due to the fact that the pion
duality interval $s_0 \approx 0.7 \, GeV^2$ 
is numerically  close to $m_{\rho}^2\approx 0.6\,GeV^2$.
Taking the lowest-order perturbative spectral density
 $\rho^{quark}(s,q^{\prime 2}=0 ,Q^2)
 = {{Q^2}/{(s+Q^2)^2}}$ and 
assuming the local quark-hadron
duality, we obtain the result
\begin{equation}
f_{\pi} F_{\gamma \gamma^* \pi^0}^{LD}(Q^2)
 = \frac1{\pi } \int_0^{s_0}  
\rho^{quark}(s,0,Q^2) \, ds
=  \frac1{\pi (1+Q^2/s_0)} 
\label{eq:FLDgg}
\end{equation}
coinciding, for $s_0=4 \pi^2 f_{\pi}^2 \approx 0.67 \, GeV^2$
 with the BL-interpolation formula.

{\it Lessons.} 
$1)$  CZ sum rule is an unreliable source
of information about the pion DA; $2)$ Pion DA is narrow;  
 $3)$ Since the diagrams 
 for the nucleon DA's  have the structure as in the pion case,
we should expect that the nucleon DA's are also close to asymptotic
and , hence, the two-gluon-exchange hard terms 
are very small for accessible $Q^2$ and $t$.
Thus, it is very important to get the estimates
for the soft contributions to the virtual Compton
scattering amplitude (see, $e.g.,$ refs. \cite{crs},  
where the high-$t$
real Compton scattering on the pion was considered).

\section{Small-$t$, large-$Q^2$ limit of VCS: a new pQCD area}

Recently,  X. Ji \cite{ji} suggested
to use the deeply virtual Compton scattering (DVCS) to 
get information about 
some parton distribution functions inaccessible 
in standard inclusive measurements.
He also  emphasized that  the DVCS amplitude 
has a scaling   behavior  
in the region  of small $t$ and fixed $x_{Bj}$
which  makes it a very interesting object 
on its own ground.

{\it  Double distributions.} 
In  the scaling limit, 
the square of the proton mass $m_p^2=p^2$ 
can be neglected compared to
the  virtuality $Q^2 \equiv -q^2$ of the initial 
photon and the energy invariant $p \cdot q \equiv m_p \nu$. 
Thus,   we set  $p^2=0$ and, for small $t$,
 we also have  $r^2 = 0$.
Then    the  requirement
$p'^2 \equiv (p-r)^2=p^2$  reduces to the condition 
$p\cdot r = 0$  which can be satisfied only
if  $r$ is  proportional to $p$: $r= \zeta  p$,
where $\zeta$ coincides with 
the Bjorken variable $x_{Bj} \equiv Q^2/2(p \cdot q)$, 
  $0 \leq x_{Bj}  \leq 1$. 
Naturally, the light-like 
limit of 4-momenta $p$, $r$ is more convenient to visualize 
in a frame where the initial proton is 
moving fast, rather than in its rest frame.

Though the momenta
$p$ and $r$ are proportional to each other,
one should make a clear distinction between them
since  $p$ and $r$  specify 
the momentum flow in two different   channels.
 Since the initial quark  momentum 
originates both from $p$ and $r$, 
 we  write it as  
$xp +y r$. 
In more formal terms, 
the relevant light-cone matrix elements
are parameterized as  
\begin{eqnarray} 
&& \hspace{-6mm} \langle p-r\, | \, \bar \psi_a(0) \hat z 
E(0,z;A)  \psi_a(z) \, | \, p \rangle |_{z^2=0} 
 =  \bar u(p-r)  \hat z 
u(p) \label{eq:vec}  \\ && \hspace{-6mm} \int_0^1   \int_0^1  \, 
  \left ( e^{-ix(pz)-iy(r z)}F_a(x,y) \right.
- \left.  e^{ix(pz)-i\bar y(r z)}F_{\bar a}(x,y)
\right )
 \, \theta( x+y \leq 1) dy \, dx ,
\nonumber 
\end{eqnarray} 
$etc.,$ where $\hat z \equiv \gamma_{\mu} z^{\mu}$ 
and $\bar u(p-r), u(p)$ are the Dirac spinors for the nucleon.

Taking the  limit $r =0$ gives  the matrix
element defining the parton distribution functions $f_a(x)$,
$f_{\bar a}(x)$.
This  leads to  the reduction formula:
\begin{equation}
\int_0^{1-x} \, F_a(x,y)\, dy=  f_a(x) .
\label{eq:redf}
\end{equation}

{\it  Asymmetric distribution  functions.}
Since $r = \zeta p$,  the variable $y$ appears 
in eq.(\ref{eq:vec}) only in 
$x+y\zeta \equiv X$ and $x- \bar y\zeta \equiv  X - \zeta$ 
combinations,
where $X$ and  $(X - \zeta)$ are   the total fractions 
of the initial hadron momentum $p$ carried by the  quarks.
Integrating the  double   distribution $F(X-y \zeta,y)$ 
over $y$ we  get the asymmetric distribution function
\begin{equation}
{\cal F}_{\zeta}^a (X) =  \int_0^{{\rm min} \{ X/\zeta, 
\bar X / \bar \zeta \}} F_a(X-y \zeta,y) \, dy, \label{eq:asdf}
\end{equation}
where $\bar \zeta \equiv 1- \zeta$.
Since $\zeta \leq 1$ and  $x+y \leq 1$, 
the variable $X$ satisfies a natural
constraint $0\leq X \leq 1$.
In the region $X > \zeta$ (Fig.4$a$),
the initial quark momentum $Xp$ 
is larger than the momentum transfer $r = \zeta p$,
and we can treat ${\cal F}_{\zeta}^a  (X)$ 
as  a generalization of the 
usual  distribution function $f_a(x)$. 
In this case, 
 the  quark  goes out of 
the hadron with a positive fraction  $Xp$  
of the original hadron momentum
and then comes  back into the hadron with a changed 
(but still positive) fraction  $(X - \zeta)p$.
The Bjorken ratio   $\zeta$  specifies
 the momentum asymmetry  of 
the matrix element. Hence, one 
deals now  with a family of 
asymmetric distribution functions  ${\cal F}_{\zeta}^a (X)$ 
whose shape changes when $\zeta$ is changed.
The  basic distinction  between 
the  double  distributions $F(x,y)$ 
and the asymmetric  distribution functions 
${\cal F}_{\zeta} (X)$ is that   the former  
do not depend on the momentum asymmetry  parameter
$\zeta$, while the latter are explicitly labelled by it.
\begin{figure}[htb]
\mbox{
   \epsfxsize=5.5cm
 \epsfysize=2.5cm
 \hspace{2.1cm}  
 \epsffile{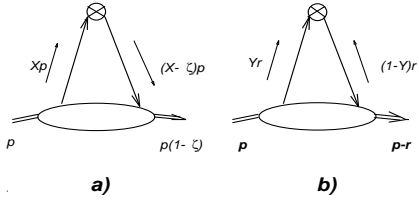} 
 }
{\caption{\label{fig:4}
Momentum flow corresponding to the  asymmetric  distribution 
function in two regimes:
{\it a) } $X \geq \zeta$ and {\it b)}
 $X \equiv Y \zeta \leq \zeta $.
}
}
\end{figure}
When $\zeta  \to 0$, the  limiting curve 
for ${\cal F}_{ \zeta}(X)$ reproduces the 
usual distribution function:
 \begin{equation}
 {\cal F}^a_{\zeta=0} \, (X) =   f_a(X) \  . \label{eq:Fzeta0}
\end{equation}
Another region is $X < \zeta$ (Fig.4$b$), 
in which  the ``returning''  quark has
a negative fraction $(X- \zeta)$ of the light-cone momentum $p$.
Hence, it is more appropriate to  treat it  as an antiquark
going out of the hadron and 
propagating  together with the original quark.
Writing $X$ as $X = Y \zeta$, we see that 
the  quarks  carry now 
positive fractions $Y \zeta p \equiv Y r$ and
  $\bar Y r  \equiv (1-Y)r $ of 
the momentum transfer $r$, and  
 the
asymmetric distribution function 
in the region $X= Y \zeta < \zeta$
looks like a distribution amplitude 
$\Psi_{\zeta}(Y)$ for a  $\bar q q $ 
state with the  total momentum $r= \zeta p$: 
\begin{equation}
\Psi_{\zeta}(Y) =  \int_0^Y F((Y-y) \zeta , y ) \, dy . 
\label{eq:Psi}
\end{equation}

{\it Leading-order contribution.}
Using the parameterization for the matrix elements
given above, we get a parton-type representation
for the handbag contribution at $t=0$:
$$
T^{\mu \nu}_{symm} (p,q,r) =  
\left (g^{\mu \nu} -\frac1{p \cdot q } 
(p^{\mu}q^{\nu} +p^{\nu}q^{\mu}) \right ) \, 
\sum_a 
e_a^2\, \sqrt{1- \zeta} \ ( T_V^a(\zeta ) + 
T_V^{\bar a}(\zeta ) ) ,
$$
where only the $\{\mu \leftrightarrow \nu \}$-symmetric  
part is shown explicitly 
and 
$T_V^a(\zeta )$ is the 
invariant amplitude 
depending on  the scaling variable $\zeta $:
\begin{equation} 
T_V^a(\zeta ) =  \int_0^{1}  \left
 ( \frac1{X-\zeta +i\epsilon}
+ \frac1{X} \right ) {\cal F}_{\zeta}^a (X) \,   dX  \, .
\label{eq:tv}  
\end{equation} 
The term  containing $1/(X-\zeta +i\epsilon)$ 
generates the imaginary part:
\begin{equation} 
- \frac1{\pi}\, {\rm Im} \, T_V^a(\zeta ) = 
{\cal F}^a_{\zeta} \, (\zeta)\,  .
\label{eq:imtv}
\end{equation}

Though 
${\cal F}_{\zeta = 0}^a(X) = f_a(X)$,  
 in the general
case when $\zeta \neq 0$, these two functions differ.
Furthermore,  the imaginary part 
appears for $X= \zeta$, $i.e.,$ in a
highly asymmetric  configuration in which the second quark
carries a vanishing fraction
of the original hadron momentum, in contrast 
to the usual distribution
$f_a(\zeta)$ which corresponds to a symmetric 
 configuration  with  the final quark 
having   the momentum equal to that of  the initial one.
A characteristic feature of the asymmetric distribution
functions ${\cal F}_{\zeta}^a(X)$ is that 
they    rapidly  vary  in the region $X \lapprox \, \zeta$ 
and  vanish for $X=0$. 
However,  the limiting curve ${\cal F}_{\zeta=0}(X)$
does  not necessarily vanish for $X=0$,
$i.e.,$ the limits $\zeta \to 0$ and $X \to 0$ do not commute.
For this reason,  if $\zeta$ is small, the substitution  of 
${\cal F}_{\zeta}^a(X)$ by $f_a(X)$ may be a good approximation
for all $X$-values except for the region $X \lapprox \, \zeta$, and 
 it is not 
clear {\it a priori}  how close are the functions
 ${\cal F}_{\zeta}^a(\zeta)$ and $ f_a(\zeta)$.

{ \it Evolution of the double distributions.} 
The purely scaling behavior of the DVCS amplitude
is violated by the logarithmic $Q^2$-dependence of $F_{NS}(x,y;Q^2)$  
 governed by the
evolution equation
\begin{equation}
Q \frac{d}{d Q} F_{NS}(x,y;Q^2) = 
\int_0^1 d \xi \int_0^1  R_{NS} (x,y; \xi, \eta;g) 
F_{NS}( \xi, \eta;Q^2) d \eta  
\label{eq:nfwdev} 
\end{equation}
(the flavor-nonsinglet (NS) component is 
taken for simplicity).
Since  integration over $y$ converts $F_{NS} (x,y)$ 
into the parton distribution function $f_{NS} (x)$,
whose evolution is governed by the GLAPD  
equation \cite{gl,ap,d},  our kernel 
 has  the property
\begin{equation}
\int_0^ {1-x}  R_{NS} (x,y; \xi, \eta;g) d y =
\frac1{\xi} P_{NS} (x/\xi).
\label{eq:rtop}
\end{equation}
For a similar reason, integrating $R_{NS}(x,y; \xi, \eta;g)$
over $x$ one should get  the evolution   kernel $V(y,\eta;g)$
\cite{blpi79,pl80}
for the pion distribution amplitude 
\begin{equation}
\int_0^{1-y}   R_{NS}(x,y; \xi, \eta;g) d x = V(y,\eta;g).
\label{eq:rtov}
\end{equation}
In the formal $Q^2 \to \infty$
limit,  
$F(x,y; Q^2\to \infty)  \sim \delta(x) y \bar y,$
$i.e.,$ in each of its variables $x,y$, the double distribution tends
to  the characteristic asymptotic form:
$\delta(x)$ is specific for the distribution functions,
while  the $y \bar y$-form  is  
the asymptotic shape   for the lowest-twist two-body 
distribution amplitudes \cite{pl80,blpi79}.

{ \it  Evolution of asymmetric distribution functions.}
As a result,  the evolution of the asymmetric distribution functions
${\cal F}_{\zeta}^a(X)$ proceeds in the following way.
Due to the GLAP-type evolution, 
the momenta of the partons decrease and distributions
become peaked in the regions of smaller
and smaller $X$. However, when the parton momentum
degrades  to values smaller than the momentum transfer
$r = \zeta p$, the further evolution is like 
that for a distribution amplitude:
it tends to make the distribution symmetric with respect to
the central point $X= \zeta/2$ of the $(0, \zeta)$ segment.

{\it Conclusions.}  DVCS opens a new class of scaling phenomena
characterized by absolutely new nonperturbative functions
describing  the structure of the proton.
The continuous electron beam accelerators like TJNAF
and ELFE may be an ideal place to study DVCS \cite{afanas}.
The asymmetric distributions can also be studied in the
processes  of large-$Q^2$ meson electroproduction
\cite{gluon}, $etc.$

{\it Acknowledgement.} This work was supported by the US Department of 
Energy under contract DE-AC05-84ER40150.

\end{document}